\documentclass{jpp}

\usepackage[utf8]{inputenc} 
\usepackage[T1]{fontenc}      
\usepackage[english]{babel}
\usepackage{graphicx}
\usepackage[a4paper]{geometry}
\usepackage{makeidx}
\usepackage{amsmath,amsfonts,amssymb}
\usepackage{fancyhdr}
\usepackage{moreverb}
\usepackage{here}
\usepackage{gensymb}
\usepackage{natbib}
\bibpunct{(}{)}{;}{a}{}{,}

\usepackage[colorinlistoftodos]{todonotes}

\usepackage{changes}
\definechangesauthor[color=blue]{FC}

\newcommand{\reff}{Fig.\ref}
\newcommand{\refs}{Sec.\ref}
\newcommand{\refe}{Eq.\ref}
\newcommand{\reft}{Tab.\ref}

\shorttitle{Interplay between KHI and LHDI}
\shortauthor{J. Dargent, F. Lavorenti, F. Califano, P. Henri, F. Pucci and S. S. Cerri}

\title{Interplay between Kelvin-Helmholtz and Lower-Hybrid Drift instabilities}

\author{J\'er\'emy Dargent\aff{1} \corresp{\email{jeremy.dargent@df.unipi.it}}, Federico Lavorenti\aff{1,2}, Francesco Califano\aff{1}, Pierre Henri\aff{2,3}, Francesco Pucci\aff{4} and Silvio S. Cerri\aff{5} }

\affiliation{
\aff{1} Dipartimento di Fisica "E. Fermi", Università di Pisa, Pisa, Italy
\aff{2} LPC2E, CNRS, Orl\'eans, France
\aff{3} Laboratoire Lagrange, CNRS, Observatoire de la Cote d’Azur, Universit\'e Cote d’Azur, Nice, France
\aff{4} Centre for Mathematical Plasma Astrophysics, Department of Mathematics, KU Leuven, Belgium
\aff{5} Department of Astrophysical Sciences, Princeton University, Princeton, NJ 08544, USA
}

\begin{document}

\maketitle

\begin{abstract}
Boundary layers in space and astrophysical plasmas are the location of complex dynamics where different mechanisms coexist and compete  eventually leading to plasma mixing.
In this work, we present fully kinetic Particle-In-Cell simulations of different boundary layers characterized by the following main ingredients:
a velocity shear, a density gradient and a magnetic gradient localized at the same position.
In particular, the presence of a density gradient drives the development of the lower hybrid drift instability (LHDI), which competes with the Kelvin-Helmholtz instability (KHI) in the development of the boundary layer. 
Depending on the density gradient, the LHDI can even dominate the dynamics of the layer. 
Because these two instabilities grow on different spatial and temporal scales, when the LHDI develops faster than the KHI an inverse cascade is generated, at least in 2D. This inverse cascade, starting at the LHDI kinetic scales, generates structures at scale lengths at which the KHI would typically develop.
When that is the case, those structures can suppress the KHI itself because they significantly affect the underlying velocity shear gradient.
We conclude that depending on the density gradient, the velocity jump and the width of the boundary layer, the LHDI in its nonlinear phase can become the primary instability for plasma mixing. 
These numerical simulations show that the LHDI is likely to be a dominant process at the magnetopause of Mercury.
These results are expected to be of direct impact to the interpretation of the forthcoming BepiColombo observations.
\end{abstract}

\section{Introduction}

Boundary layers spontaneously emerge in many space and astrophysical plasmas, typically in the presence of two interacting different plasma environments.
That is the case of the solar wind interacting with self-generated or induced magnetospheres of various Solar System's objects, of the interaction between the local interstellar medium and the heliopause or of astrophysical jets interacting with the surrounding environment, just to mention a few. 
Such layers typically exhibit large-scale variations of magnetic, density and/or velocity fields.
Because those gradients represent a source of free energy for a variety of plasma instabilities, their evolution feeds back into the global evolution of such systems.
A typical situation of interest for the above-mentioned cases is provided by the boundary layer forming between two different magnetized plasma flows, as it occurs at planetary magnetopauses \citep[e.g.,][]{Fujimoto1998,Hasegawa2003,MastersPSS2012,Cerri2013,Haaland2014,LiljebladJGRA2015,Cerri2018,Malara2018}. 
In particular, such shear flows can be unstable to Kelvin-Helmholtz instability (KHI), developing the characteristic fully rolled-up vortex structures \citep[e.g.,][]{Nakamura2005,Henri2013,Faganello2017}.
Signatures of possible KHI structures have been indeed observed at several planetary magnetpauses~\citep[see e.g.,][]{Hasegawa2004,SundbergJGRA2012,DelamereJGRA2013,Paral2013,LiljebladJGRA2014,GershmanJGRA2015}.
These vortices may in turn feed secondary instabilities, developing on the shoulder of the primary KHI.
A typical example is provided by the vortex-induced magnetic reconnection in various forms \citep[e.g.,][]{Nakamura2008,Faganello2009,Nakamura2013,Nakamura2014,Fadanelli2018}, or by the development of pressure anisotropies able to trigger kinetic instabilities \citep[e.g.,][]{Decamillis2016}.
Moreover, when the two plasma flows have different densities, the Rayleigh-Taylor instability (RTI) may be triggered by the large-scale vortex motion \citep{Matsumoto2004,Faganello2008,Faganello2008b}. 
At fluid scales, the RTI can also emerge as a primary instability because of the gravity acceleration since the magnetosheath is denser than the magnetosphere.
However, since the gravity acceleration is small and because the magnetic field tends to stabilize, the RTI is usually much less considered with respect the KHI, except than in some special cases \citep{Guglielmi2010}.

At kinetic scales, a Lower-Hybrid Drift instability (LHDI) may also emerge from the density gradient itself \citep{Gary1993,Daughton2003,Daughton2004}.
The LHDI is often observed in both spacecraft data \citep{Mozer2011,Norgren2012,Graham2014,Graham2017b,Yoo2018,Yoo2019} and laboratory experiments \citep{Carter2001,Carter2002,Yoo2014,Yoo2017} and has been especially studied in the context of magnetic reconnection \citep{Lapenta2002,Pritchett2012,Roytershteyn2012,Roytershteyn2013,Price2016,Le2017,Le2018}.
As a consequence, the KHI and the LHDI are expected to compete and  interact in a quite complex way since their fastest growing mode (FGM) grows at different scale lengths and with different growth rates.
The dynamics arising from such a competition and its nonlinear development is the main focus of this paper. 

The KHI is a fluid-scale instability growing in a sheared flow.
Its growth rate is controlled by the velocity shear and the corresponding gradient scale length \citep{Chandrasekhar1961,Miura1982,Faganello2017}.
On the other hand, the LHDI is a kinetic scale instability.
This instability is driven by the coupling of ion thermal gyration with the free energy provided by the ion density gradient drift velocity.
At scales where ions are demagnetized but electrons still frozen-in, this free energy efficiently feeds Lower-Hybrid waves through an inverse ion damping that makes these waves unstable \citep{Gary1993}.
The growth rate is controlled by the frequency ratio between the electron plasma frequency and the cyclotron frequency, the plasma beta and the density gradient.
In this work, we will mainly focus on the latter.

Due to the different typical scale length at which the KHI and the LHDI develop, the interplay between these instabilities has not yet been observed.
The KHI including a density gradient has been investigated either by adopting a MagnetoHydroDynamic model \citep{Takagi2006,Faganello2008,Matsumoto2010,Leroy2017} or by adopting a hybrid kinetic model but neglecting electron inertia \citep{Gingell2015}.
In this case, however, the LHDI can not grow since the FGM scales as  $k_{FGM} \propto (mi/me)^{1/2}$ \citep{Gary1990} and stabilizes more and more when the density gradient scale length becomes larger than a few ion inertial lengths \citep{Gary1993}. 
There also exists some fully kinetic simulations theoretically able to develop LHDI but none of them has actually shown evidence of it, either because the layer is too large \citep{Matsumoto2006,Umeda2010}, either because the KHI have been excited to grow quicker \citep{Matsumoto2010,Umeda2014}.
As a consequence, no simulation of KHI ever reported the growth of LHDI in space plasma.
On the other hand, kinetic simulations of KHI revealed another kinetic effect: the dawn-dusk asymmetry, which impact the growth of the KHI depending on the sign of $\mathbf{B} \cdot \mathbf{\Omega}$, where $ \mathbf{\Omega}$ is the vorticity at the layer \citep{Nakamura2010,Henri2013,Paral2013}.

In this work, in order to investigate the interplay between the KHI and LHDI, we make use of a kinetic model of a boundary layer characterized by the presence of (i) a velocity shear that would be unstable to the KHI and (ii) a density gradient that would be unstable to the LHDI. 
We take the scale lengths of variation of the main fields (densities, magnetic field and velocities) of the order of the ion inertial length and vary their respective values. 
These configurations allow us to study the relative impact of each instability on the layer and on the plasma mixing.

The present paper is organized as follows.
\refs{sec:param} describes the numerical model used in this paper. 
\refs{sec:res} presents the results, and is split into the different phases of the simulations.
\refs{ssec:LHDI} shows the linear growth of the LHDI, when applicable, and the comparison of our numerical results with linear theory.
\refs{ssec:nonlinear} contains the nonlinear phase of the LHDI and  describes the presence of an inverse cascade of energy responsible for the formation of large scale structures.
In some regime of parameters, those structures are shown to interfere with the growth of the KHI that develops on larger time scales. 
This feature, together with the factors playing a role in the competition between the nonlinear phase of the LHDI and the linear growth of the KHI are studied in \refs{ssec:KHI}.
In \refs{sec:dis}, we discuss the results of this paper and the inherent limitations of the used model.
Finally, in \refs{sec:con}, we present a summary of this study, together with a description of possible consequences for the dynamics of planetary magnetospheres, such as the hermean magnetopause.

\section{Numerical setup}
\label{sec:param}

We present four fully kinetic simulations in a two-dimensional (2D) cartesian geometry using the Particle-In-Cell (PIC) code SMILEI \citep{Derouillat2017}.
The simulations model the boundary between two plasmas characterized by different densities and separated by a velocity shear.
Simulations are performed in the reference frame where the low-density plasma medium is at rest.

The data presented are normalized using ion scale quantities.
The magnetic field and density are normalized to arbitrary value $B_0$ and $n_0$, respectively.
We choose $B_0$ and $n_0$ such that the density and magnetic field are equal to one on the flowing side of the layer (in our simulations: the right side).
The masses and charges are normalized to the proton mass $m_p$ and charge $e$,
time is normalized to the inverse of the proton gyrofrequency $\omega_{ci}^{-1}=m_p/eB_0$ and
length to the proton inertial length $\delta_i=c/\omega_{pi}$, where $c$ is the speed of light and $ \omega_{pi}=\sqrt{n_0e^2/m_p\epsilon_0}$ is the proton plasma frequency.
Velocities are normalized to the ions' Alfv\'en velocity $v_{Al} = \delta_i \omega_{ci}$.

All simulations are initialized with a single layer where density, velocity (directed along the $y$-direction) and magnetic field (directed along the $z$-direction) vary along the $x$ direction.
This layer is contained in the $(x,y)$ plane in a 2D domain of size $(x_{max},y_{max})=(68,136)~\delta_i$.
There are $n_x=n_y=2720$ cells in the $x$ and $y$ directions, corresponding to a grid resolution of $\Delta_x = 0.025~\delta_i$ and $\Delta_y = 0.05~\delta_i$.

The ion and electron distribution functions are initially composed by $50$ macro-particles per cell loaded using Maxwellian distributions.
Plasma moments and electromagnetic forces are calculated using second order interpolation. 
The time step is calculated using a Courant-Friedrichs-Lewy (CFL) condition which in our simulations turns out to be $\Delta_t=8.4\cdot 10^{-4} ~\omega_{ci}^{-1}$, and the total simulation time is $400~\omega_{ci}^{-1}$.
The boundary conditions are periodic in the $y$ direction and reflective in the $x$ direction for both particles and fields.

The initial density profile is given by:
\begin{eqnarray}
	n_i(x,y) &=& \frac{1}{n_r} \left[ 1 + \frac{n_r-1}{2} \left( 1+ \tanh \left( \frac{x-x_0}{L}  \right) \right) \right] + n_{curr}
	\label{eq:ni}
\end{eqnarray}
 where $n_r=n_{a}/n_{b}$ is the density ratio across the current sheet, $L$ the initial characteristic width of the layer, fixed at $L=1$ in this work, and $n_{curr}=\mathbf{J}^2/(e^2n_0v_{Al}^2)$ a small correction to allow ions to carry part of the current in order to avoid inconsistencies (negative electron density in case of very strong magnetic field gradient).
 The subscript $a$ and $b$ stands for the asymptotic values on both sides of the layer.
The values of $n_r$ are listed in \reft{tab:asym}. 
The ions velocity is initialized as:
\begin{equation}
	\mathbf{v}_i(x,y) = \left[\frac{\Delta v_{shear}}{2}  \left( 1+ \tanh \left( \frac{x-x_0}{L} \right)  \right)
	\right] \mathbf{e}_y
	~+~ \mathbf{v}_{curr}
	\label{eq:vi}
\end{equation}
where $\Delta v_{shear} = |v_{a} - v_{b}|$ is the velocity difference across the shear layer and $\mathbf{v}_{curr}=\mathbf{J}/(en_i)$ a small correction to allow ions to carry part of the current, in order to avoid inconsistencies, as consistent with the density correction.
The values of $\Delta v_{shear}$ are listed in \reft{tab:asym}. 
The initial magnetic field profile is given by: 
\begin{eqnarray}
	\mathbf{B}(x,y) &=& \frac{1}{B_r} \left[ 1 + \frac{B_r-1}{2} \left(1+ \tanh \left( \frac{x-x_0}{L} \right) \right) \right] \mathbf{e}_z
	\label{eq:B}
\end{eqnarray}
where $B_r=|B_{a}/B_{b}|$ is the magnetic asymptotic field ratio across the current sheet.
The values of $B_r$ are listed in \reft{tab:asym}.
From the curl of $\mathbf{B}$ we calculate the current $\mathbf{J}$ since in our model the displacement current is neglected. 
Finally, the electric field is initially set to:
\begin{equation}
	\mathbf{E} =-\mathbf{v}_i \times \mathbf{B} + \frac{1}{en_i} \mathbf{J} \times \mathbf{B}
	\label{eq:E}
\end{equation}
which correspond to an Ohm's law where the initial electron pressure gradient term has been neglected. 
The $\nabla(P_e)/en_e$ term in the generalized Ohm's law is by far the smallest contribution to the electric field. To ease the implementation of the initial conditions, we have neglected the electron pressure gradient term, and later checked and confirmed that  this term can be neglected.
However, note that this simplified equation does not perfectly match the Ohm's law within the boundary, as we consider a charge separation ($n_i \ne n_e$) associated to the electric field gradient \citep{Pritchett1984}.
Finally, electron density $n_e = n_i - \nabla \cdot \mathbf{E}$ and mean velocity $\mathbf{v}_e = (\mathbf{J}-n_i\mathbf{v}_i)/n_e$ consistently follow from $\mathbf{J}$ and Maxwell equations.

The total scalar pressure $P=P_i+P_e$ is determined in order to preserve pressure balance. The electron to ion temperature ratio is taken constant and equal to $\theta=T_e/T_i=0.2$.
The plasma $\beta= 2P/B^2$  is set equal to 1 in plasma "a" (left side).
A reduced mass ratio $m_i/m_e = 25$ is used for computational reasons. 
The consequences of using a reduced mass ratio is explained in details in \refs{sec:dis}.
We fix $\omega_{pe}/\omega_{ce}=4$. 
To optimize the layer stability, we make use of finite Larmor radius correction in the simulation setup, as described in \citet{Cerri2014}.

\begin{table}
	\begin{center}
   \begin{tabular}{ | l || c | c | c | c |  }
     \hline
     Simulation    & 1 & 2 & 3 & 4 \\ \hline
     $B_r$ &  0.5 & 0.5 & 0.5 & 0.5 \\ \hline
     $n_r$ & 10 & 10 &  5  & 1 \\
     \hline
     $\Delta v_{shear}$ & 0 & 0.5 &  0.5  & 0.5 \\
     \hline
   \end{tabular}
 	\end{center}
   \caption{
   Magnetic field ratio ($B_r$), density ratio ($n_r$) and velocity difference ($\Delta v_{shear}$) values characterizing the discontinuity between the two adjacent plasmas, for the 4 simulations.
   Simulation 2 parameters correspond to parameters consistent with Mercury's magnetopause \citep{Slavin2008}.
   }
   \label{tab:asym}
\end{table}

\section{Numerical results}
\label{sec:res}

The simulations are characterized by three main phases.
In the first one, a Lower Hybrid Drift instability (LHDI) develops very fast in all simulations except in simulation 4 where no density gradient is present.
In the second phase, the LHDI saturates and enters the nonlinear stage, which is characterized in simulations 1 and 2 by the growth of large scale finger-like structures.
Finally, in the last phase, the Kelvin-Helmholtz instability (KHI) may develop depending whether or not the nonlinear LHDI structures grew fast enough (i.e., to a sufficiently large amplitude) to destroy the coherence of the shear flow.
In order to better illustrate the different phases, movies of the out-of-plane magnetic field time evolution are available for simulations 2, 3 and 4 as supplementary material.
In the following, we discuss how the system evolves in each of these phases for each simulation.

\subsection{The linear LHDI phase}
\label{ssec:LHDI}

\begin{figure}
    \centering
    \includegraphics[width=\linewidth]{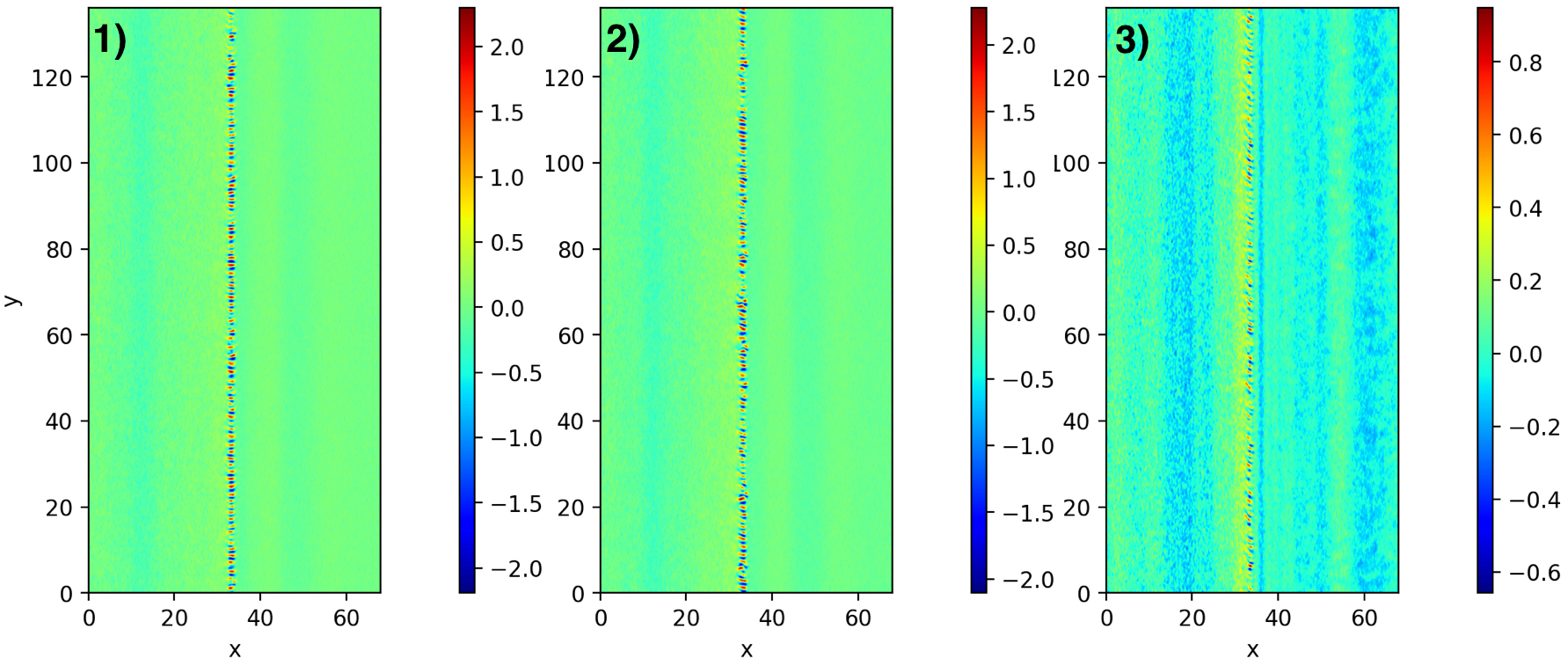}
    \caption{From left to right panel: in-plane electron velocity component $v_x$ at $t=7.5$ for simulations 1, 2 and 3, respectively.
    Simulation 4 is not plotted here because in the absence of a density gradient, the LHDI does not develop, and consequently there are no visible fluctuation at the layer this early in the simulation. 
    } 
    \label{fig:Vex7}
\end{figure}

At the initial times of all simulations (except for simulation 4), we observe growing fluctuations along the layer at $x \simeq 34$.
In particular, in \reff{fig:Vex7} we show the in-plane electron velocity component $v_x$ for simulations 1, 2 and 3, where the fluctuations length scale is observed at kinetic scale.
Furthermore, the amplitude of fluctuations is bigger in simulations 1 and 2 than in simulation 3 since the instability depends on the density gradient (see \reft{tab:asym}).
Furthermore, a wave front  propagating in the $x$ direction is clearly visible as a consequence of the absence of a kinetic equilibrium in the layer's initialization and our choice to neglect the pressure gradient term in \refe{eq:E}.
However, this artificial by-product of the initialization has not a significant impact on the evolution of the simulation, as typically the case for kinetic simulation.
Those waves are especially clear in simulation 3 since the color scale amplitude has a reduced range with respect to simulations 1 and 2. 

To identify the instability as a LHDI, we compare our results with the corresponding linear theory. 
We solve the linear Vlasov equation under some simplified assumptions to compute the growth rate of the LHDI \citep{Gary1978,Gary1983,Gary1993}.
The simplifications are made in order to carry out the analytical model and are the following:
density and temperature gradients much larger than ion gyro-radius, uniform out-of-plane magnetic field and a low plasma beta, $\beta= 2P/B^2<<1$.
Our method consists in integrating the linearized Vlasov equation along unperturbed orbits, thus finding an expression for the dielectric function, and finally finding the zeros of this function numerically.

In our case, we have used a dieletric function $\varepsilon(k, \omega)$ of the form \citep{Gary1979,Sgro1989,Gary1993} :
\begin{eqnarray}\label{eq:Vlasov_dispersion}
    \begin{aligned}
    \varepsilon(k, \omega) &= 1 + \sum_{j} K_j(k, \omega) \\
    K_j(k, \omega) &=\frac{A_j^2}{k^2}\bigg(1-[\omega-kv_{nj}] e^{-k^2 \rho^2_{Lj}}\sum_{m=-\infty}^{+\infty}\frac{I_m(k^2 \rho^2_{Lj})}{\omega + m\omega_{cj}}+ \\ 
    &+\frac{\omega v_{nj}}{k v^2_{th,j} }e^{-k^2 \rho^2_{Lj}}\sum_{m=-\infty}^{+\infty} \frac{m \omega_{cj} I_m(k^2 \rho^2_{Lj})}{\omega + m\omega_{cj}}\bigg)-\\
    &-k^3\rho^2_{Lj} v_{Tj} e^{-k^2 \rho^2_{Lj}}\sum_{m=-\infty}^{+\infty} \frac{I_m(k^2 \rho^2_{Lj})- I'_m(k^2 \rho^2_{Lj})}{\omega + m\omega_{cj}}\\
    \end{aligned}
\end{eqnarray}
\noindent
where $k$ and $\omega$ are the wave vector and frequency, respectively, and $K_j$ is the dielectric susceptibility of the species $j$ (the sum over $j$ running on all the species, $j=(i,e)$ in our case).
$I_m$ are the modified Bessel functions of order m. 
Moreover, $v_{th,j}= \sqrt{T_j/m_j}$  and $\rho_{Lj}= \sqrt{m_j T_j}/q_j B$ are the thermal velocity and the Larmor radius of the species $j$, respectively, and $B$ is the magnetic field modulus.
Finally $m_j$ and $q_j$ are the mass and charge of species $j$, respectively, and $\omega_{cj}=q_jB/m_j$ is the gyrofrequency of the population $j$.
The drift velocities associated with the density ($n_j$) and temperature ($T_j$) gradients are defined as $v_{Fj}=\epsilon_{_F} \rho_{Lj} v_{th,j}$, where $F_j=(n_j,T_j)$ and $\epsilon_{_F} = ({1}/{F_j(x)}) d F_j(x) / dx$.
We also define $A_j= \omega_{pj}/v_{th,j}$.
All the above formulas are in ion units ($\delta_i$ and $\omega_{ci}^{-1}$).

As already said, this method assumes a gradient configuration with a density and temperature gradient length scale ($L_F\sim\epsilon_F^{-1}$) much larger than ion gyro-radius ($\rho_{Li}/L_F\ll1$), a uniform out-of-plane magnetic field and a low plasma beta, $\beta = 2P/B^2<<1$.
Although those assumptions do not strictly hold in our simulations, the neglected effects would tend to slow down the process, so that we can consider the analytical theory as an upper limit for the growth rates and a useful reference for a comparison with simulation results in the linear regime \citep{Davidson1977}. 
Indeed, since $v_{nj}$ and $v_{_{Tj}}$ are proportional to the density and temperature gradients, the LHDI growth rate is expected to be larger for steeper gradients of $n$ and $T$. 
In the configuration adopted for the linear theory, we shall define the inverse gradient length $\epsilon_{_{T,n}}$ as the one computed at $x \simeq x_0$, i.e. at the initial location of the shear layer (see \refs{sec:param}).

The linear theory assumes a uniform magnetic field $B$ (and thus a uniform Larmor radius $\rho_{_{Li}}$), while in the simulation the magnetic field is sheared.
Therefore, we must fix a characteristic (mean) value of $\rho_{Li} = \sqrt{\beta_i m_i/2 e n_i} \equiv \sqrt{\beta_i /2 n_i}$ in normalized units. 
We take it as the local value at the center of the magnetic gradient, i.e. at $x=x_0$, and report the results in \reft{tab:grad}.

\begin{table}
	\begin{center}
   \begin{tabular}{ | l || c | c | c |  }
     \hline
     Simulation    & 1-2 & 3 & 4 \\ \hline
     $\epsilon_n $ & 0.82 & 0.67 & 0 \\ \hline
     $\epsilon_T $ & -0.7  & -0.48  & -0.27 \\ \hline
     $\rho_{Li}  $ & 2.18 & 1.73 & 1.42 \\
     \hline
   \end{tabular}
 	\end{center}
   \caption{Gradients and Larmor radius in the different simulations.
   }
   \label{tab:grad}
\end{table}

\begin{figure}
    \centering
    \includegraphics[width=\linewidth]{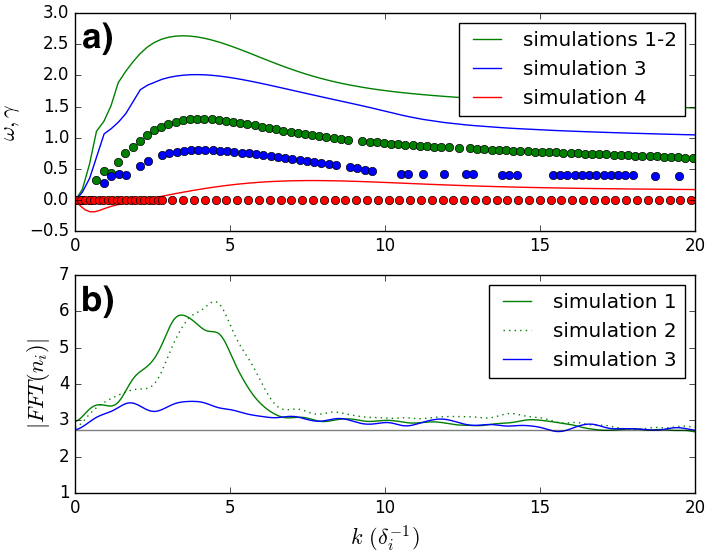}
    \caption{a) Theoretical dispersion relation of the LHDI branch obtained from \refe{eq:Vlasov_dispersion} with $m_i/m_e=25$, and for the set of parameters given in \reft{tab:grad}.
    The dotted lines represent the growth rate and the solid lines represent the real frequencies.
    b) Fourier transform along y of the density field at x=34 and t=5 for simulation 1, 2 and 3.
    }
    \label{fig:LHDIrate}
\end{figure}

We have computed the dispersion relation using the parameters listed in \reft{tab:grad}.
The corresponding results, associated to the four kinetic simulations, are reported in \reff{fig:LHDIrate}.

The main result emerging from \reff{fig:LHDIrate}$a$ is that the LHDI is expected to grow on  ion kinetic scales in simulations 1-2 and 3, while in simulation 4 the system is stable with respect to the LHDI.
The LHDI fastest growing mode is 
$k_{FGM} = 3.9$ for simulations 1-2 and 3, while the growth rate is $\gamma_{LHDI}=1.3$ in simulation 1-2 and $\gamma_{LHDI}=0.8$ in simulation 3 because of the different density gradients.
In \reff{fig:LHDIrate}$b$ we show the density spectrum observed in our simulations at $t=5$.
The peak around the FGM is in agreement with linear theory.
On the other hand, simulation 4 does not develop any instability at such early times.
As expected, the instability grows faster in simulation 1 and 2 than in simulation 3. 
We therefore conclude that the instability observed in the early stage of simulations 1, 2 and 3 corresponds to a LHDI.

The growth rate in simulations is estimated by fitting the temporal evolution of the Fourier transform of $n_i$ (see \reff{fig:LHDIrate}.$b$) for $k=4$ (the fastest growing mode) during the linear phase of LHDI with an exponential function $Ae^{t \gamma_{LHDI}}$.
Quantitatively, the measured growth rates in simulations 1, 2 an 3 turn out to be smaller, by a factor of three, than the ones estimated by linear theory (see \refe{eq:Vlasov_dispersion}).
Such a discrepancy is explained by the inherent limitations of the theoretical model \citep{Davidson1977}, which considers a smooth density gradient, a uniform magnetic field and a low plasma $\beta$.
Finally, it is worth noticing that the presence of a velocity shear does not affect too much the instability, as shown in \reff{fig:LHDIrate}$b$, where simulations 1, with no velocity shear, and 2, with a velocity shear, behave similarly.

\subsection{The nonlinear LHDI phase}
\label{ssec:nonlinear}

\begin{figure}
    \centering
    \includegraphics[width=\linewidth]{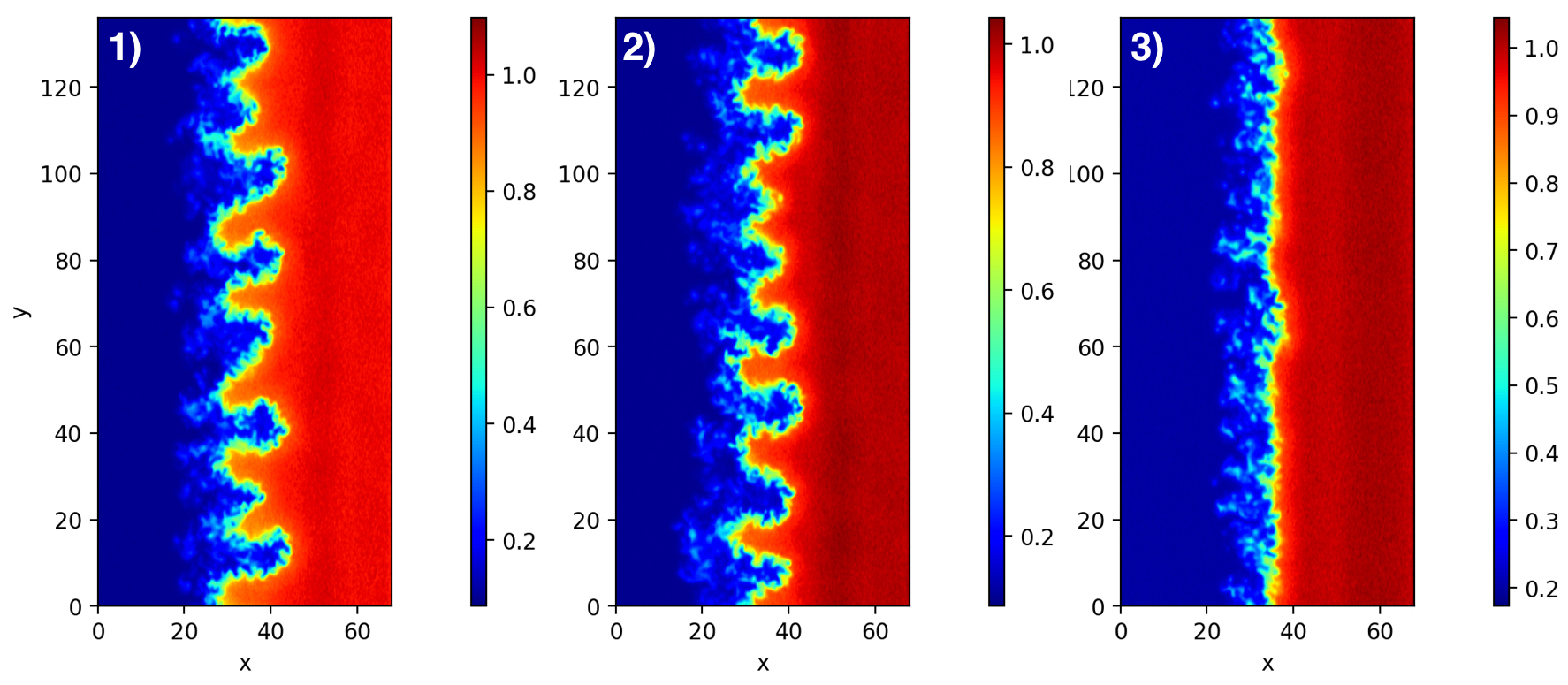}
    \caption{Ion density at $t=150$ for simulations 1, 2 and 3.}
    \label{fig:fingers}
\end{figure}

The LHDI ends its linear growth phase long before the KHI starts to develop, thus entering its nonlinear stage. 
In simulations 1 and 2 such transition to the non linear regime occurs at around $t_{NL}\sim8$, eventually leading to an effective inverse cascade with the formation of many fluid scale structures.
This is shown in \reff{fig:fingers} where we draw the density contours for simulations 1, 2 and 3 in the nonlinear stage of the LHDI. 
Furthermore, in simulations 1 and 2 we observe the development of a new process associated to the inverse cascade that produces elongated finger-like structures entering one into the other plasma. 
This is not observed in simulation 3 at similar times since the inverse cascade is less developed due to a slower growth of the LHDI.
Such phenomenon of finger-like structures generation is similar with what has been previously observed in kinetic simulations~\citep{Brackbill1984,Gary1990,Singh1998}. 

As stated above, since in simulation 3 the LHDI is less efficient, the inverse cascade occurs later and is much less intense.
Even though some finger-like features start to develop at later times, their growth is slower in simulation 3, on a time scale comparable with the growth of the KHI. 
Indeed, in this case the competition between the growth of nonlinear LHDI structures and linear KHI actually prevents the finger-like structures to grow, eventually suppressing them completely once the KH vortices form.
This last stage where KHI can develop will be further discussed in \refs{ssec:KHI}.

\begin{figure}
    \centering
    \includegraphics[width=1\linewidth]{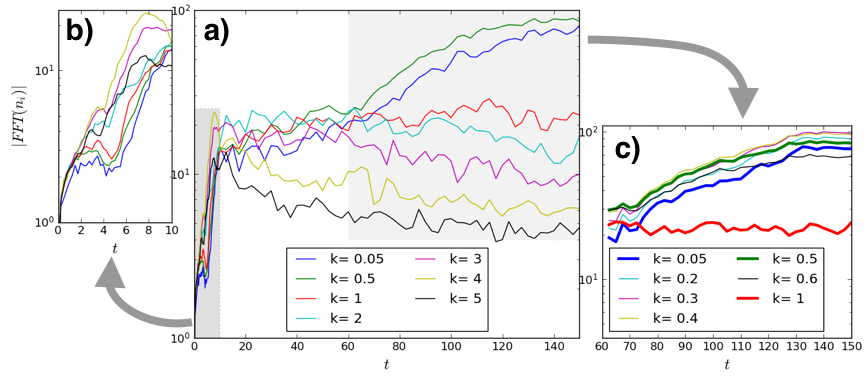}
    \caption{
    Time evolution of Fourier coefficients modulus of ion density fluctuations , $|\widehat{\delta n}_{i,k}(x)|=|\mathrm{FFT_y}(n_i)|$, for some given $k$ at $x=33.25$ from simulation 1.
    Panel $a$ and $b$ share the same k, while panel $c$ plot other k.
    The k shared by all panel are plotted with thick lines in panel $c$.
    }
    \label{fig:evolmodes}
\end{figure}

In \reff{fig:evolmodes}$a$ we show the time evolution of the modules of different Fourier coefficients for density of simulation 1.
\reff{fig:evolmodes}$b$ shows a zoom of \reff{fig:evolmodes}$a$ at early times, when the energy growth is driven by the linear development of the LHDI.
\reff{fig:evolmodes}$c$ shows a zoom at late times of \reff{fig:evolmodes}$a$, although plotting other $k$.
\reff{fig:evolmodes}$c$ shows small k ($k \leqslant 1$) in order to show the evolution of the inverse cascade at large scales.

As discussed in \refs{ssec:LHDI} the FGM of the LHDI is at about  $k=4$ (yellow curve).
However, in the nonlinear phase after $t_{NL} \sim 8$, we observe 
that this mode stops growing and its associated energy starts to decrease. This transition marks the saturation phase of the LHDI and the beginning of the inverse cascade phase which feeds the growth of lower $k$'s modes.
Indeed, after $t_{NL}$ all wave numbers larger than $k=4$ (yellow and black curves) decrease while smaller wave numbers continue to grow. 
After some time, the low $k$ modes begin to decrease one after the other, except for $k<1$ which continues to grow even faster. 
On longer times, we observe in \reff{fig:evolmodes}$c$ that the inverse cascade continues, with the energy evolving towards ever larger scales (i.e. smaller $k$).

To calculate a characteristic time-scale associated with the inverse cascade process and the corresponding growth of the fluid-scale structures, we measure the position $x_0(y)$ of the center of the layer at any value of $y$. 
To get $x_0(y)$, we fit each cut along $x$ of the density with an hyperbolic tangent and take the inflection point of the fit as $x_0(y)$. 
The time evolution of the standard deviation $\sigma_0=\sqrt{<x_0^2>_y}$ gives an estimation of the growth of the finger-like structures.

\begin{figure}
    \centering
    \includegraphics[width=1\linewidth]{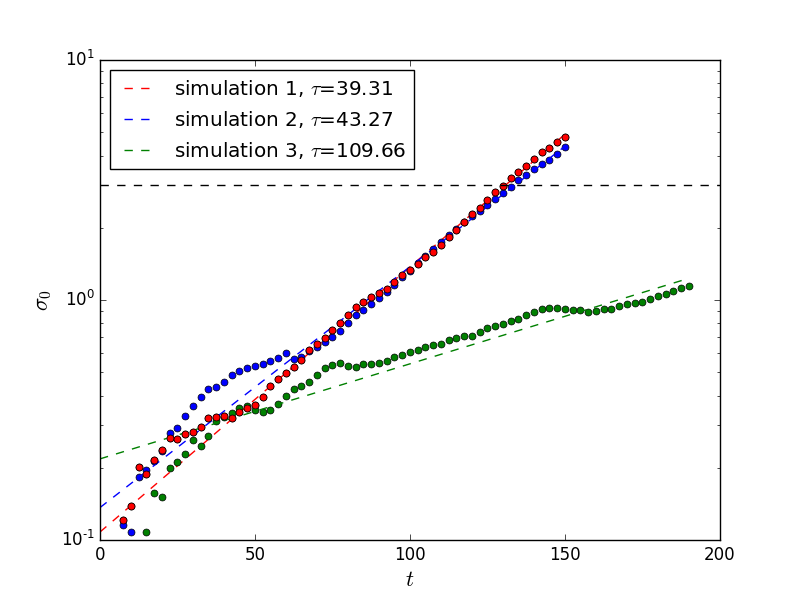}
    \caption{Time evolution of $\sigma_0=\sqrt{<x_0^2>_y}$ from simulations (dots) and corresponding exponential fit (dashed lines). Different colors represent different simulations (see legend).
    The characteristic time $\tau_{NL}$ obtained from those fits is given in the legend.
    The horizontal black dashed line gives the value of $\sigma_0\sim 3$ that the KHI reaches in its nonlinear stage at $t \sim 300$, in simulation 4.
    }
    \label{fig:stdev}
\end{figure}
The time evolution of $\sigma_0$ is shown in \reff{fig:stdev} for simulations 1, 2 and 3.
The growth of $\sigma_0$ turns out to be fitted quite well by an exponential function of the form $Ae^{t/\tau_{NL}}$, especially for simulations 1 and 2. 
We therefore determine a characteristic growth time $\tau_{NL}$ of the nonlinear LHDI structures from the exponential fit of the curves in \reff{fig:stdev}.
Note that the exponential fit match very well an exponential growth for simulation 1 where no velocity shear flow is present.
Despite the similarities between simulations 1 and 2, we observe that in the early phase simulation 2 does not match very well an exponential behavior and that the observed characteristic time of the nonlinear LHDI  structures is a bit longer than the case with no velocity shear.
Thus we conclude that the velocity shear has a small impact on the early growth of the large-scale fluid structures.
The exponential fit match even less well for simulation 3, which we attribute to the competition between the nonlinear LHDI, weakened by a smaller density gradient, and the KHI.

\subsection{The KHI phase}
\label{ssec:KHI}

\begin{figure}
    \centering
    \includegraphics[width=\linewidth]{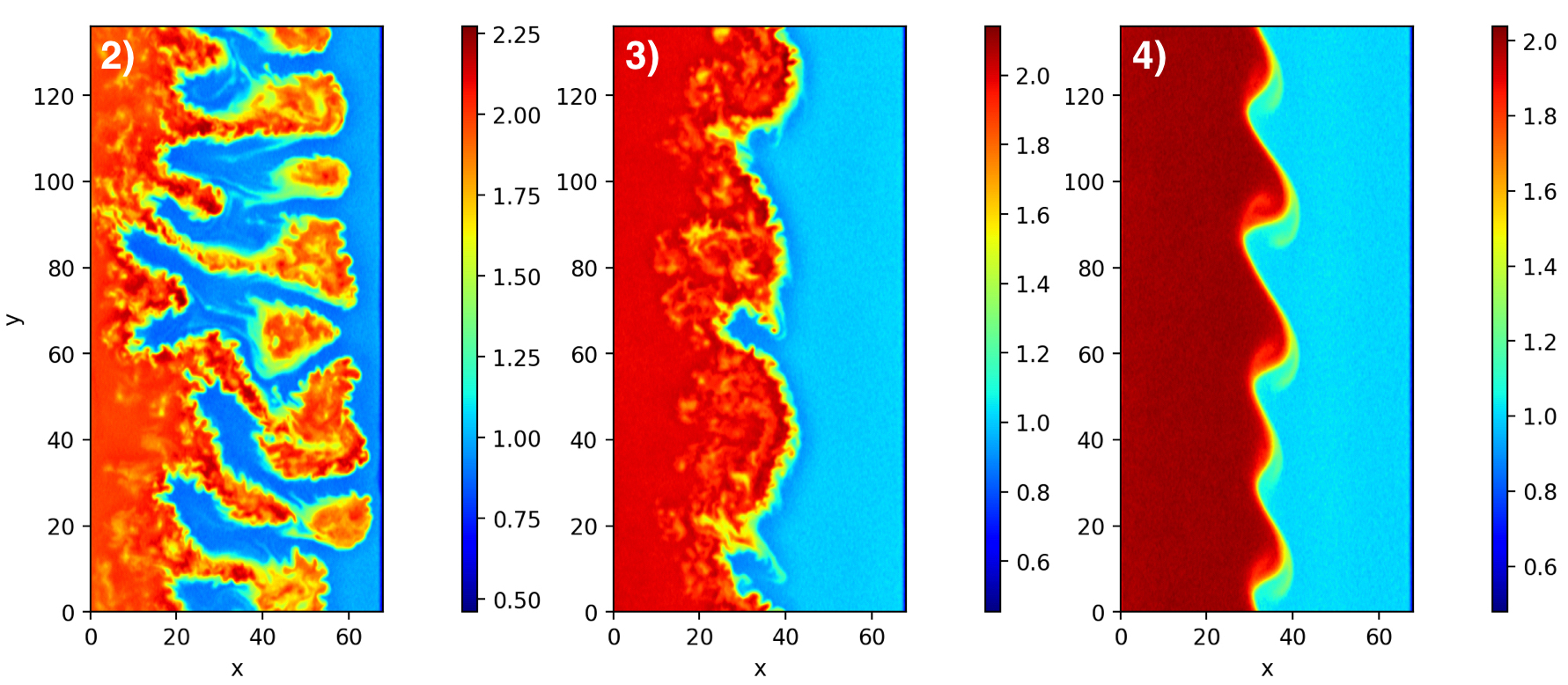}
    \caption{Magnetic field along $z$ at $t=350~\omega_{ci}^{-1}$ for simulations 2, 3 and 4.}
    \label{fig:KH}
\end{figure}

Since a velocity shear is present in simulations~2, 3, and 4 the onset of a KHI could be expected.
\reff{fig:KH} shows the out-of-plane magnetic field at $t=350~\omega_{ci}^{-1}$ for the three simulations with a velocity shear.

In simulation 4, where the LHDI does not develop, a series of KHI vortices forms.
In simulation~3, despite the patchy layer produced by LHDI, the KHI dominate the large scale nonlinear dynamics to form KHI vortices.
Note that the fastest growing mode is different between simulations~3 and~4, despite having the same initial velocity shear and layer width.
This is likely due to the widening of the initial shear layer induced by the LHDI in simulation 3.
Finally, simulation 2 does not develop KH vortices. 
Nevertheless, the dynamics observed in simulation 2 is very similar to that of simulation 1 (which does not have a velocity shear--not plotted here), with only one difference that finger-like structures are drifting along y in the former, while they do not in the latter.
The nonlinear finger-like structures developed by the LHDI grow too quickly and reach the size of the expected KHI (black dashed line in \reff{fig:stdev}) vortices long before the KHI could actually emerge.
Thus, the layers coherence is long gone at the time were we would expect KHI.
In simulation 2, the KHI has been killed by the nonlinear LHDI before it can develop.

Now we try to estimate the growth rate of the fastest growing mode of the KHI, i.e. $\gamma_{KH}$, in all simulations with a velocity shear.
In order to do that, we use the result of \citet{Michalke1964} (table 1 in the article): the fastest growing mode of the KHI over a layer with a hyperbolic tangent shape, $\tanh(x/L)$, is given by $k_{KH}L=0.44$, and $\gamma_{KH} = 0.095  \Delta u /L$. 
Such result is obtained considering an incompressible fluid, in the absence of a magnetic field and with no density asymmetry.
The incompressibility approximation roughly holds because the fast magnetosonic Mach number $M_f= u/\sqrt{c_A^2 + c_s^2}$ ($c_A$ the Alfv\'en velocity and $c_s$ the sound velocity) is always smaller than 0.05 on both sides.
We can however expect that minor compressibiliy effects would tend to just slightly reduce the growth rate of the KHI in the above incompressible limit.
Then the presence of a magnetic field here also plays no relevant role because it is perpendicular to the propagation plane.
Note however that the presence of a magnetic field gradient can also generate instabilities \citep{Huba1980b}, but we neglect this term as we don't see any $\nabla B$-drift induced instability in our simulations.
For what concerns the homogeneous density approximation, in general we can expect that a density asymmetry $n_r$ would indeed alter the results of \citet{Michalke1964}.
However, following \citet{Chandrasekhar1961}, we do expect that the growth rate would be modified as follows:

\begin{equation}
    \gamma_{KH} \propto \frac{\sqrt{n_r}}{1 + n_{r}} 
    \label{eq:gKH}
\end{equation}

\begin{table}
	\begin{center}
   \begin{tabular}{ | l || c | c | c |  }
     \hline
     Simulation    & 2 & 3 & 4 \\ \hline
     $\gamma_{KH}$ & 0.013 & 0.018 & 0.024 \\ \hline
     $\tau_{KH}  $ & 77 & 57 & 42 \\ 
     \hline
   \end{tabular}
 	\end{center}
   \caption{Theoretical growth rates $\gamma_{KH}$ of the KHI for a layer width $L$ picked at $t=10$.
   $\tau_{KH}=1/\gamma_{KH}$ is the characteristic time of the KHI growth.}
   \label{tab:KH}
\end{table}

In \reft{tab:KH}, we have computed the growth rate for the KHI in the three simulations with the parameter $L$ computed at $t=10.0$, when the linear growth of the LHDI is over. 
The width $L$ of the layer is calculated by fitting the profile of $v_y$ averaged along $y$  with \refe{eq:vi}, where $x_0$ and $L$ are left as free parameters.
At this time the layer width is more or less the same for all the simulations, corresponding to $L\sim2$.
However, the layer width $L$ keeps growing slowly but continuously in simulations 2 and 3 because of the nonlinear LHDI.
Thus, in practice, the KHI growth rate is decreasing with time in those simulations.

$\tau_{KH}$ is calculated for simulation 4 with the method used in \reff{fig:stdev} (i.e. by fitting the curve of $\sigma_0(t)$ during its exponential growh, given that $\sigma_0$ is proportional to the KH wave amplitude) gives us a result of $\tau_{KH} \approx 49$. 
Our theoretical $\tau_{KH}$ is actually in good agreement with the observed one (42 vs 49), the slight discrepancy can be due to the error associated with $L$, and the wave vector of maximum growth matches with the one observed in the simulation $k=0.23$.
These results help us to clarify what's happening in the simulations.

In simulation~2 and~3 the LHDI grows much faster than the KHI, so we expect the LHDI to start the inverse-cascade process well before the KHI begins.
Indeed we see that the cascade begins at $t\sim10$ and reaches fluid scales ($k\sim1$) at $t\sim20$.
After that time the competition between the nonlinear LHDI and the KHI begins.
In simulation~2, $\tau_{NL}<<\tau_{KH}$, so the nonlinear LHDI totally dominate the evolution and the KHI has no opportunity to grow.
In simulation~3, $\tau_{NL} \sim 2 \tau_{KH}$ so even if the nonlinear LHDI grows significantly, the KHI can still dominate.
This case is interesting as both instabilities manage to develops enough to both impact the layer structure.
In simulation~4 the LHDI is suppressed because there isn't any density gradient. 
So we see the formation of KHI vortices with a growth rate comparable to the one predicted by the theory.

\section{Discussions}
\label{sec:dis}

In this work we have shown that, despite the difference in characteristic scale length and growth rate values between the KHI and the LHDI, in the presence of a relatively strong density gradient the LHDI  not only contributes to the dynamics but even dominates at large scales.
The validity of these results and their application to space plasmas are discussed in this section.

In conditions of a boundary layer with both a velocity shear and a density gradient, the KHI and LHDI might compete. 
This competition relies on the linear time scales at which the instabilities develop and on the nonlinear time scale at which the layer diffuses and relaxes. 
All that can be summarized by three basic factors: the velocity shear amplitude, the density gradient amplitude and the layer's width.
When the importance of the density gradient dominates with respect to the velocity shear, then the structures generated during the nonlinear phase of the LHDI broaden, eventually smoothing the average velocity shear layer, thus slowing down (e.g. simulation~3) or even preventing (e.g. simulation~2) the KHI growth. 
A broadening of the layer then tends to decrease the growth rates of both KHI and LHDI. 
Moreover, for layers larger than a few ion inertial lengths,  the LHDI growth rate rapidly decreases and eventually becomes stable.
For example, with simulation 1 parameters, the LHDI becomes stable for $L \gtrsim 8$.
Therefore, when the velocity shear and the density gradient are characterized by nearly the same scale length, the layer width controls whether or not the LHDI  develops. 
This is the reason why most of the planetary magnetopauses might be stable to the LHDI, as they are usually larger than a few ion inertial lengths, and why previous fully kinetic simulations of KHI at the Earth's magnetopause did not see it \citep{Matsumoto2006,Umeda2010,Nakamura2014,Nakamura2017}. 
On the contrary, the hermean magnetopause, characterized by a much smaller length scale of the boundary layer width, is potentially thin enough for the LHDI to develop and drive the dynamics eventually up to inhibiting the KHI development.
Such a study of KHI in the fully kinetic regime is therefore especially relevant for Mercury.
As a matter of fact, simulation 2 which was designed with parameters expected in the magnetosphere of Mercury, shows that LHDI dominates the dynamics of such a strongly inhomogenous shear layer.

The main topic of this study is the competition between the LHDI and the KHI in an inhomogeneous boundary layer and its nonlinear saturation. 
In this last phase, the mechanism behind the inverse cascade of the LHDI is out of the scope of this paper. 
However, it is worth reminding some previous results regarding the nonlinear phase of the LHDI concerning the inverse cascade \citep{Davidson1978,Huba1978,Gary1979,Drake1984,Brackbill1984,Shapiro1994}.
On the one hand, the theory of \citet{Drake1984}, which relies on mode coupling to short-wavelength damped modes, works quite well to explain the inverse cascade observed in this work. 
In particular, the temporal evolution of the spectra associated with the inverse cascade observed in our simulations is thoroughly described by \citet{Drake1984}.
On the other hand, the modulational instability theory developed by \citet{Shapiro1994} might explain the exponential growth observed in the nonlinear phase of the LHDI (see \reff{fig:stdev}). Moreover, \citet{Shapiro1994} predicted the formation of large-scale structures similar to those observed in our simulation at later times (see \reff{fig:KH}.$1$).
These two explanations might be not exclusive.

For practical reasons linked to the full PIC modeling, our simulations use (i) a reduced mass ratio (ii) and a reduced geometry (2D). 
Thus, the scales separation is much smaller than for a realistic mass ratio and it affects the LHDI development.
Typically, the fastest growing mode of the LHDI in our simulations is for $k_{FGM}\sim 4$, while for a realistic mass ratio it would been $k_{FGM}\sim 40$ \citep{Gary1993}.
\citet{Drake1984} argued that a realistic mass ratio will support the nonlinear evolution of the LHDI due to two reasons.
First, the ratio of the rate of change of the magnetic field energy to particle drift energy scales as $m_i/m_e$ in a finite $\beta$ plasma \citep{Drake1981}, so much more energy will be available to supply the inverse cascade.
Second, the number of unstable modes scales as $\sqrt{m_i/m_e}$ \citep{Huba1980}.
So, for realistic values, we expect a much broader spectrum of unstable modes to be excited. 
For these reasons, we expect the LHDI nonlinear phase to develop faster for a realistic mass ratio as compared to our simulations.
To support our claim, we observe that the growth of the nonlinear LHDI structures is faster than ours in the paper of \citet{Gary1990}, where the mass ratio is $m_i/m_e=100$, despite a slightly smoother density gradient.
Future studies will look at the impact of the mass ratio for realistic magnetopause parameters.

As this work has been performed in a reduced 2D configuration, understanding the changes that would arise for a 3D configuration to the linear and nonlinear development of the LHDI is a future necessary step.
First of all, in 3D the instability is no more confined to the plane perpendicular to B, so we expect growing modes with $\mathbf{k}\cdot \mathbf{B}\neq0$. 
The nonlinear relaxation in 3D could lead to the formation of elongated cigar-shaped structures \citep{Shapiro1994}, to the acceleration of electrons parallel to the magnetic field \citep{Singh1998,Bingham2002}, and to mode conversion towards different kinds of waves such as, e.g., whistlers mode \citep{Camporeale2012}. 
How such effects would affect the competition and interaction between the LHDI and the KH is unclear and will be investigated in the future.

The work described in this paper highlights the importance of processes other than KHI to generate large scale structures responsible for plasma mixing along the magnetopause. 
In regions characterized by a strong density inhomogeneities, such as the magnetopause of Mercury, the LHDI provides another efficient mechanism for plasma mixing, together with the KHI reported by MESSENGER observations \citep{Slavin2008}.
Some differences between the structures generated by KHI and LHDI should help us to identify them. 
Typically, during its late nonlinear phase, the KHI generates a diffuse layer \citep{Matsumoto2010}, while the LHDI generates large scale finger-like structures that do not evolve into a diffusive layer \citep{Brackbill1984,Gary1990,Singh1998,Bingham2002}.
Future studies should be dedicated to the interplay between KHI and LHDI structures in observations in order to understand the actual role of the LHDI in an hermean-like magnetosphere. 
This requires observations at the lower hybrid frequency, not available from MESSENGER data, but planned with the BepiColombo mission, especially with the PWI consortium onboard the Mio (MMO) spacecraft \citep{Kasaba2010}.

\section{Conclusion}
\label{sec:con}

In this paper, we have performed several 2D-3V kinetic simulations to study the linear and nonlinear evolution of a magnetized shear flow separating plasmas of different densities and magnetic field intensity.
This study shows that the large scale structures that emerge from the nonlinear phase of the LHDI can interfere with the KHI development if the nonlinear LHDI driven dynamics modifies the velocity shear layer. 
In a layer subjected to both a velocity shear and a density gradient, we have shown that, at early time and within the parameters range used in this study, a LHDI develops first.
Then, the LHDI enters a nonlinear phase and energy starts to cascade from small to large scales. Depending on the growth rate of those structures, the system can become totally dominated by the nonlinear LHDI and the KHI won't be able to develop.
On the contrary, if the nonlinear LHDI growth rate is smaller than the KHI growth rate the KHI dominates and leads to the formation of vortices.
However, those vortices are already partially mixed due to the simultaneous, even if less efficient, development of the LHDI at smaller scales.

The range of parameters used in this study encompasses those expected in the magnetosphere of Mercury (e.g. simulation 2).
We therefore expect that these kind of structures could be detected by future observations provided by the BepiColombo space mission. 

\section*{Acknowledgments}

This project (JD, FC) has received funding from the European Union's Horizon 2020 research and innovation programme under grant agreement No 776262 (AIDA).
We acknowledge ISCRA for awarding us access to the supercomputer Marconi at
CINECA, Italy, where the calculations were performed. We thank M. Guarrasi (CINECA) for useful discussion about code implementation on Marconi.
The work at LPC2E/CNRS was supported by CNES and by ANR under the financial agreement ANR-15-CE31-0009-01.
The work by F. Pucci has been supported by Fonds Wetenschappelijk Onderzoek - Vlaanderen (FWO) 
through the postdoctoral fellowship 12X0319N.
S.S.C. is supported by the National Aeronautics and Space Administration under Grant No.~NNX16AK09G issued through the Heliophysics Supporting Research Program.

\bibliographystyle{jpp}
\bibliography{biblio}
\end{document}